# Arrays of Nano-Electromechanical Biosensors Functionalized by Microcontact Printing

*Sven Salomon,[1,2] Thierry Leïchlé,\*[1,2] Denis Dezest,[1,2] Florent Seichepine,[1,2] Samuel Guillon,[1,2] Christophe Thibault,[1,2] Christophe Vieu,[1,2] and Liviu Nicu[1,2]*

[1]CNRS, LAAS, 7 avenue du colonel Roche, F-31400 Toulouse, France

[2]Univ de Toulouse, LAAS, F-31400 Toulouse, France

tleichle@laas.fr

ABSTRACT

The biofunctionalization of nanoelectromechanical structures is critical for the development of new classes of biosensors displaying improved performances and higher-level of integration. We propose a modified microcontact printing method for the functionalization and passivation of large arrays of nanocantilevers in a single, self-aligned step. Using fluorescence microscopy and resonant frequency measurements, we demonstrate (1) the bioactivity and the anti-fouling property of deposited antibodies and BSA molecules and (2) the preservation of the nanostructures' mechanical integrity.

KEYWORDS

Nanoelectromechanical systems, Biosensing, Biofunctionalization, Microcontact Printing, Cantilever

MANUSCRIPT TEXT

High-frequency nanoelectromechanical systems (NEMS) are attracting more and more interest as a new class of sensors and actuators for potential applications ranging from computation[1] to single (bio)molecule sensing.[2, 3] For NEMS to be considered as a viable alternative to their actual biosensing macro counterparts, they have to simultaneously meet three major requirements: high mass responsivity (MR), low minimum detectable mass (MDM) and low response time (RT).

Without any doubt, as emphasized by theoretical studies,[4] the two first specifications (high MR and low MDM) can be successfully addressed by NEMS devices. Such predictions have been already validated in case of virus sensing,[5] enumeration of DNA molecules[6] or even single molecule nanomechanical mass spectrometry.[2] Hopelessly, nanometer scale sensors have been proved, still in theory,[7, 8] to be inadequate to practical RT scales which, if confirmed as such, could definitely impede the way towards realistic biosensing applications. To prevent this from happening, one possible trade-off strategy would consist in taking advantage from considering a single NEMS device not alone but as part of a functional array of similar devices.[9] This paradigm allows, while preserving the benefits of high MR and low MDM of a single device, to use the considerably higher capture area of the NEMS array such as the RT reaches practical relevance. However, in that case, the non-reactive areas of the chip containing multiple sensors functionalized with a single type of probe molecule must be adequately coated with an anti-fouling film in order to lower the probability of adsorption of target molecules anywhere else than on sensitive areas and hence to permit their detection at ultra-low concentrations.

To address the production of massively parallel arrays of NEMS for biorecognition applications, one has to be able to perform uniform, reliable biofunctionalization of nanoscale devices at a large-scale (i.e. at the level of the entire NEMS array) while being able to block all other surfaces. Ideally, this could be achieved by using a technique enabling the deposition of the probe molecules and the anti-fouling layer in a single self-aligned printing step, thus ensuring that target molecules only react with the surface of the sensor.

Here we use an advanced microcontact printing (µCP) process[10] in order to biofunctionalize arrays of NEMS with a probe molecule on the active sensing area either prior to or together with an anti-fouling layer on the passive areas. We demonstrate the adequate functionalization/anti-fouling of arrays of freestanding nanocantilevers as dense as $10^5$ nanostructures/cm² by using both fluorescence microscopy and dynamic measurements of the structures' resonant frequency. The proper bioactivity of an antibody deposited onto the cantilevers, and the blocking property of a bovine serum albumin (BSA) layer are both assessed by incubating specific and non-specific tagged secondary antibodies followed by fluorescence imaging. Furthermore, measurement of the resonant frequency of the nanocantilevers before and after functionalization and biological recognition demonstrate that using µCP for device functionalization does not damage the nanostructures and preserves the mechanical sensing capability of our NEMS.

One of the major issues preventing the breakthrough of NEMS devices for biodetection applications remains the bio-functionalization of the sensor, i.e. the grafting of probe molecules onto its active surface.[11] While this technical difficulty can easily be circumvented when taking a bottom-up approach since the nanostructures can be functionalized before integrating them into a device, it raises tremendous constraints when considering NEMS devices obtained by means of micro- and nanofabrication techniques. Indeed, preservation of the bioactivity of the sensing layer imposes to avoid subsequent fabrication steps leading to the degradation or the alteration of the probe molecules, e.g. processes involving plasma and vacuum environments or harsh chemicals.[12] This means that surface modification is most likely to be carried out at the end of the fabrication process, i.e. after the release of the nanostructures. Hence, care must be taken when delivering probe molecules onto freestanding fragile structures to keep the integrity of the sensor.

Moreover, the functionalization of large array of NEMS devices implies that submicron patterns can be created at large-scale. Thus, the most common functionalization method relying on the immersion of the entire sample into a dedicated biochemical solution is not fulfilling the need to create localized biofunctionalized areas.[13] Lately, many techniques have been proposed to locally biofunctionalize microscale sensors, such as ink-jet printing[14] and the use of separate microcapillaries.[15] However, these approaches are either not likely to produce sub-microns patterns nor to be scaled up to create large and multiplexed arrays. Photolithographic light-directed synthesis,[16] already used for gene chip fabrication, are appropriate for large multiplexed arrays, but the diffraction limit makes it difficult to apply to nanoscale patterning. Liquid drop dispensers have also been demonstrated as useful tools for the biofunctionalization of MEMS sensors,[17] however, the capillary forces induced when retracting the tool from the surface are most likely large enough to damage nanostructures. Other technologies used to create nanoscale patterns by means of tip-based liquid or molecular transfer, such as miniaturized fountain pen[18, 19] and scanning-probes,[20] could possibly address the issues related to the biofunctionalization of fragile nanostructures, even if the creation of patterns on large areas seems technically difficult to implement. Finally, one can think of using interesting approaches relying on the use of localized electrochemical reactions induced at the surface of addressable electrodes,[21, 22] despite the problems foreseen to be raised by the high-multiplexing requirements of the electrode array in order to access individual or groups of nanostructures.

In this paper, we propose to use microcontact printing,[23] which is a highly parallel deposition technology[24] with nanoscale resolution,[25] for the localized biofunctionalization of large-scale arrays of nanostructures. µCP is a "dry" deposition technology known to keep biomolecules fully active after deposition. Our proposed method relies on a modified µCP process, where the trenches in the stamp are actually used to transfer the molecules of interest, while the base of the stamp provides mechanical support during the printing step and can additionally be used to deposit blocking molecules on the non-reactive parts of the array. The stamp is designed so that when placed on a surface, the bottom of the trenches comes into contact with that surface if a pressure is applied, thus allowing molecular transfer without inducing too much force onto the structure to functionalize.[26]

The functionalization process, illustrated in the Figure 1, includes two stages: the polydimethylsiloxane (PDMS) stamp preparation and inking and the actual printing or transfer process. First, the PDMS stamp is incubated with the probe biomolecules, thus covering the entire stamp surface (the top and bottom of the trenches). In order to remove probe molecules from the top of the trenches, i.e. where the stamp is supposed to contact the non-reactive areas of the chip to functionalize, the stamp is brought into contact with a bare glass substrate. Few contacts on clean glass slides are usually operated in order to ensure that no traces of molecules are left at the surface of the stamp. This step, so-called "cleaning step", results in the presence of biomolecules solely into the trenches of the stamp. Thus, during this cleaning step, it is crucial to make sure that the PDMS patterned trenches do not collapse in order to keep the bottom of the trenches correctly inked. Then, in order to deposit an anti-fouling layer outside the biofunctional patterns, the stamp is incubated with a surface-blocking molecule (Figure 1B). The addition of this inking step leads, after print, to the patterning of probe molecules and anti-fouling agents in a single, self-aligned step. This method is thus especially suited to the single-step functionalization and passivation of planar structures, such as membranes. Alternatively, the deposition of the anti-fouling layer can also be carried out in a batch process, after functionalizing the array of nanostructures (Figure 1A). Transfer of the molecules from the stamp onto the chip to functionalize is finally achieved by bringing the stamp into contact with the array of NEMS, after aligning features patterned on both surfaces and deforming the stamp under pressure until the bottom of the cavities gently touches the nanostructures. Deforming the stamp and inducing the collapse of the trenches to transfer biomolecules ensures that a minimum pressure is applied onto the structures and thus prevents any damage or breakage of the NEMS.

In the present work, we demonstrate the feasibility of our approach by biofunctionalizing arrays of freestanding nanocantilevers. More than $10^5$ nanocantilevers were fabricated onto 1.5 cm × 1.5 cm chips from silicon-on-insulator wafers using standard microfabrication techniques, i.e. UV stepper photolithography, silicon reactive ion etching and $SiO_2$ wet etching in buffered oxide etch (see Supporting Information). Fabricated cantilevers with typical dimensions of 8 µm in length, 2.6 µm in width and 340 nm in thickness are freestanding above 15 µm x 12.5 µm x 1 µm long, wide and deep cavities (Figure 2a). The PDMS stamp used for biofunctionalization purposes was designed to fit the pitch of the nanocantilever array and the size of the patterned trenches was adjusted to the dimensions of the cavity under each nanocantilever, i.e. 15 µm x 12.5 µm, such as to increase alignment efficiency. In that case, even with a misalignment of several microns, which is the minimum typical tolerance of alignment tools found in academic labs, the probe biomolecules are still transferred to the entire surface of the nanocantilevers (since the patterning area is larger than each cantilever). In fact the only alignment issue could originate from a discrepancy between the designed stamp and its actual size after de-molding: indeed, due to the shrinkage of the polymer during the baking step, the pattern size of fabricated stamps typically shrinks by about 1%. Even if this effect is not likely to affect printing results for most applications, the dimension compliance cannot be ensured when working with arrays of thousands of nanocantilevers spanning over few centimeters: thus even if the stamp is perfectly aligned along one side of the array, the patterns would be shifted from few lines or columns on the opposite side. Hence, to avert this shrinkage effect, we have created composite stamps consisting of patterned 1.5 cm $\times$ 1.5 cm $\times$ 400 µm PDMS films glued to 1.5 cm $\times$ 1.5 cm $\times$ 1 mm thick glass slides by placing these glass slides in a dedicated set-up during the PDMS casting onto silicon master molds (Figure S1). This fabrication trick resulted in a stamp deformation lower than 0.05%.

Alexa Fluor 660 donkey anti-goat IgG (IgG1) and albumin from bovine serum (BSA) fluorescein conjugate were printed in a single self-aligned step respectively onto and around large arrays of nanocantilevers using the printing process described in Figure 1B. For this purpose, the fabricated PDMS stamp was firstly inked with a 400 µg/mL Alexa Fluor 660 donkey anti-goat IgG in 1X Phosphate Buffered Saline (PBS) solution during 2 min, rinsed and dried. It was then inked with a 1 mg/mL BSA solution for 2 min after stamping several times onto a bare glass slide to remove all the IgG1 at the extruded surface of the stamp (i.e. not within the trenches). After rinsing and drying the stamp, it was aligned with the array of nanocantilevers and then gently brought into contact with the NEMS chip, resulting in the adhesion of the PDMS stamp at the surface of the chip and thus in the transfer of BSA molecules around the nanostructures. Finally, the transfer of donkey anti-goat IgG from the stamp to the nanocantilevers was achieved by applying a pressure onto the stamp to induce the deformation of the patterned trenches so that the PDMS would contact the cantilevers, as clearly seen in the movie M1. The pressure was applied by placing the stamp and the NEMS chip within a vice and the pressure level and the duration of the print were assessed by studying the level of fluorescence emitted by the transferred molecules (i.e. by seeking and obtaining a maximum level of fluorescence, comparable to the one resulting from printing onto a planar $SiO_2$ surface) while avoiding breaking the nanocantilevers. The as-biofunctionalized arrays of nanocantilevers were imaged by fluorescence microscopy using two filter sets dedicated to the observation of the Alexa Fluor 660 and the Fluorescein dyes (Figure 2b). As can be seen in the inset of Figure 2b, the levels of raw fluorescence indicate that indeed the IgG1 is solely patterned onto the nanocantilevers, while the entire remaining surface of the chip is covered with BSA molecules. These results indicate the proper deposition of the probe and the anti-fouling biomolecules at the desired location of the chip, without apparent breakage of the nanocantilevers.

Next, to demonstrate the bioactivity of the deposited donkey anti-goat IgG and the anti-fouling property of the BSA layer, we carried out biodetection experiments involving exposure to goat anti-mouse IgG (Specific IgG2) and donkey anti-mouse IgG (Non-specific IgG2), inducing respectively specific and non-specific interactions with the patterned IgG1. For the purpose of these experiments, the surface blocking step was carried out in a batch mode, following the process displayed in Figure 1A, rather than using the double stamp transfer. Indeed, the single and self-aligned biofunctionalization and passivation stamp process, which is perfectly suited for planar surfaces (e.g. nanomembranes or nanoelectrodes), does not enable the deposition of molecules underneath the nanocantilevers and inside the cavities: these surfaces would not be properly blocked and would thus be subject to non-specific adsorption of the incubated molecules, leading to false positive response of a mass sensor (which is our chosen detection scheme) due to the added mass at the back-side surface of the nanocantilevers. Following IgG1 print onto the cantilever top surface and a batch immobilization of BSA molecules everywhere else (i.e. additionally at the bottom of the cavities and underneath the cantilevers), a specific recognition of the printed donkey anti-goat IgG was carried out by incubating the functionalized chips in a 50 µg/mL Alexa Fluor 488 goat anti-mouse IgG solution for 2 h. Other functionalized chips were also incubated in a 50 µg/mL Alexa Fluor 488 donkey anti-mouse solution for control experiments. After incubation, the NEMS chips were thoroughly washed 10 times with 1 mL degassed PBS solution, soaked into PBS for 45 min, then rinsed 10 times with degassed DI water and finally dried with a nitrogen gun before fluorescence imaging. The use of degassed solutions for the batch deposition of the anti-fouling layer as well as for incubation with specific and non-specific IgG2 prevented bubble formation into the cavities underneath the nanocantilevers, thus allowing the entire surface of the chip to contact the solutions and avoiding accumulation of molecules at the bubble contact lines as well as inhomogeneity and poor efficiency of the rinsing steps.

The results of the incubation experiments are resumed in Figure 3. Fluorescence images in Figure 3a show the adequate immobilization of Alexa Fluor 660 donkey anti-goat IgG at the surface of the nanocantilevers that cannot be seen using the filter cube dedicated to the observation Alexa Fluor 488 labeled specific and non-specific IgG2. Non-fluorescently labeled BSA was deliberately chosen for this experiment to avoid disrupting the signal emitted by the Alexa Fluor 488 of the secondary IgG2. Figure 3b, top and bottom, shows fluorescence images obtained using the filter sets dedicated to the Alexa Fluor 488 after incubation respectively with non-specific and specific IgG2. While a slight non-specific interaction is observed all over the chip, the specific interaction can clearly be seen at the surface of the nanocantilevers, thus demonstrating the adequate passivation of the chip and the bioactivity of the deposited IgG1 at the nanocantilevers' surface. Evidence of these results is reinforced when plotting the increase of fluorescence signals averaged over the blocked and functionalized areas of a single chip before and after incubation with IgG2 (Figure 3c). Indeed, the increase of fluorescence at the surface of the nanocantilevers is meaningfully larger in the case of specific interaction with a high signal to noise ratio, while the fluorescence of the negative control is comparable to the level of non-specific adsorption observed on the BSA layers for both specific and non-specific molecules. Thus, the nanocantilevers biofunctionalized with our modified µCP method are adequately proven to work as biosensors.

Finally, measurements of the resonant frequency and the quality factor corresponding to the fundamental mode of vibration of the functionalized nanocantilevers were then conducted to address two questions: do the nanocantilevers preserve their mechanical integrity once the functionalization and the successive incubation steps have been done? Are the calculated mean added masses - relative to the resonant frequency shifts measured before functionalizing the nanocantilevers and after incubating with the specific IgG - relevant from a biological point of view?

In order to answer these questions, we introduce mechanical considerations as a basis for the interpretation of the measurements performed on the nanocantilevers in a dynamic regime. When biological molecules are added onto the surface of a cantilever, the total effective mass increases by $\Delta m$ resulting in a proportional shift of the NEMS resonant frequency $\Delta f$, given by:

$$\qquad \qquad (1)$$

where $f_0$ and $M_{eff}$ are the resonant frequency and the effective modal mass of the cantilever. Theoretically, the resonant frequency and the effective modal mass of the designed cantilevers are respectively equal to 6.87 MHz and 4.1 pg. In order to estimate a mean mass added on our cantilevers, we have used the resonant frequency $f_0$ measured after removal of the biological layers, the measured resonant frequency change, and the effective modal mass estimated from the cantilever dimensions. Experimentally, after fluorescence observation and measurement of the cantilever resonant frequency, the biological layers were completely removed by immersing the NEMS chips into a Piranha bath during 10 min.

The question of the mechanical integrity was thoroughly investigated by performing systematic resonant frequency and quality factor measurements on 16 nanocantilevers randomly chosen among thousands of structures of a single functionalized array. The main and first conclusion is that the nanostructures undergo the different biological treatments without damage, i.e. these structures are not broken and are still vibrating. Moreover, the resonant frequency and corresponding quality factor values decrease in a reproducible way, as expected, after loading of the biological matter (Figure 4). Indeed, the quality factor decrease is attributed to a damping effect induced by presence of the biological layer on top the surface of the nanocantilevers, while the resonant frequency systematic drop is due to total mass added during the biological protocol sequences. The latter assumption neglects the surface stress effects that the biomolecules may exert onto the dynamic behavior of the nanocantilevers. This hypothesis is supported by the fact that the thickness of the transferred material by µCP onto the active area of one nanocantilever represents only a tiny fraction of its overall thickness.[27]

The question related to the biological relevance of the mean added mass value is rather delicate to address from a quantitative point of view. Table 1 provides the mean measured resonant frequency shift, the corresponding mean added mass per unit area, and the relative mean variation of measured quality factor values for 16 nanocantilevers after biofunctionalization with the IgG1, incubation with the BSA protein and interaction with the specific IgG2. The resonant frequencies and quality factors of bare nanocantilevers considered as reference values are obtained after removing the biological layers. The standard deviation values are indicated in the brackets along with each of the measured and calculated parameters indicated above.

A qualitative analysis of the Table 1 confirms that mass-loaded cantilevers exhibit a simultaneous decrease of resonant frequency and of the quality factor. Moreover, the mass addition value of 18,6 fg/µm$^2$ is significantly higher than the one previously reported after solely printing the IgG1, i.e. 11.7 fg/µm$^2$,[26] which is in accordance with the extra mass due to the BSA molecules adsorbed underneath the cantilevers and the goat anti-mouse IgG specifically interacting with the bioactive layer. Ideal comparison between the resonant frequency of a same cantilever measured after each biological protocol step, i.e. after biofunctionalization, surface blocking and incubation, would provide data directly proving the actual biosensing capability of our functionalized nanocantilever resonators. However, this approach is banned because of the modus operandi used to perform the measurements. Due to the vacuum conditions imposed by the Fabry-Perot experimental set-up, it would be risky to assume that the biological molecules grafted onto the nanocantilevers are not irreversibly denatured or damaged once submitted to such extreme conditions after the dynamic characterizations. To prevent arguable conclusions in case of biological protocols taking place onto the same chip that necessarily undergoes back-and-forth steps between vacuum and liquid media, we deliberately decided to carry out the entire biological protocol before performing the resonant frequency measurements, and then to remove the final biological arrangement to obtain the initial resonant frequency. For this reason, we are currently unable to provide direct comparison of resonant frequency shifts induced by the incubation of specific and non-specific interacting molecules, since the added mass due to the printed IgG1 and incubated BSA cannot be measured directly before the revelation steps. One way to prevent such a critical experimental configuration would be to perform all biological reactions and subsequent measurements in liquid media and in real-time, as classically done with tools such as quartz microbalance.[28] However, it has become obvious that the nanocantilever geometry is fundamentally inadequate to dynamic sensing in liquid due to the dramatic impact of the surrounding medium onto the quality factor and as a result onto the minimum detectable mass by such structures. With the advent of new NEMS geometries more adapted to biosensing in liquid media, we are foreseeing to provide soon biofunctionalized integrated NEMS sensors by means of the patterning approach presented in this paper. Still, with these preliminary results, supported by fluorescence microscopy, we can safely conclude that the biofunctionalization technique presented here is a first resolute step towards setting-up bioNEMS as the ultimate mechanical systems for complex and multiplexed biosensing operations.

In this paper, we have presented a technique for the biofunctionalization and the surface blocking of a large array of nanostructures based on a single, self-aligned µCP process. Our method relies on the use of a PDMS stamp displaying patterns inked with probe and anti-fouling molecules at the bottom and the top of the trenches, respectively. While the anti-fouling layer is transferred upon direct contact of the stamp with the surface of the chip, probe biomolecules are delivered by deforming the trenches so that the PDMS bottom patterns come into contact with the nanostructures. As a proof of concept, we have functionalized arrays of more than $10^5$ nanocantilevers/cm$^2$ with donkey anti-goat IgG, while blocking the remaining chip surface with BSA. The bioactivity an anti-fouling properties of the deposited molecules were assessed by incubating the NEMS chips with goat anti-mouse IgG and donkey anti-mouse IgG, inducing specific and non-specific recognition of the printed IgG, following by fluorescence observations. Next, measurement of the nanocantilever resonant frequency demonstrated that the nanostructures retain their mechanical integrity and thus their sensing capability, after the printing and the incubation steps. We believe that the technique presented in this paper answers the biofunctionalization challenge of nanostructures for the development of future generation of functional bioNEMS.

ACKNOWLEDGMENT

The French National Agency for Research (program ANR/PNANO 2008, project NEMSPIEZO 'ANR-08-NANO-015') is gratefully acknowledged for financial support.

Supporting Information.

Details of the materials and methods concerning (1) the fabrication of the arrays of silicon nanocantilevers, (2) the fabrication of the PDMS stamps, (3) the printing process, (4) the molecular incubation and imaging protocols and (4) the measurement of the nanocantilever resonant frequency. Figure S1 showing schematics of the stamp fabrication using a dedicated stainless steel set-up. Movie M1 displaying a real-time video of the printing process.

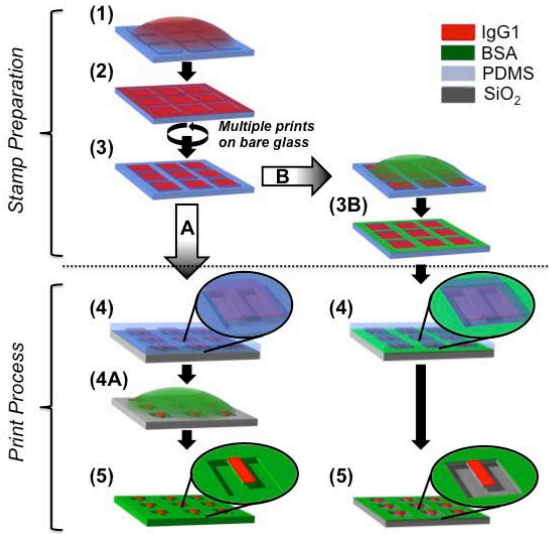

Figure 1. Schematics of the NEMS functionalization process using a PDMS stamp; two approaches can be used: A. Surface functionalization by microcontact printing followed by blocking the remaining

surfaces with a batch process. B. Surface functionalization and passivation both carried out by microcontact printing. (1) Inking the stamp with the desired molecules (IgG1). (2) Washing and drying the stamp. (3) Cleaning outside the stamp grooves, i.e. removing the molecules from the stamp base, by carrying multiple prints. (3B) Inking the stamp with the anti-fouling molecules (BSA). (4) Printing after aligning the stamp and the chip. (4A) Incubating the chip with the anti-fouling molecule (BSA). (5) Result of the NEMS biofunctionalization process.

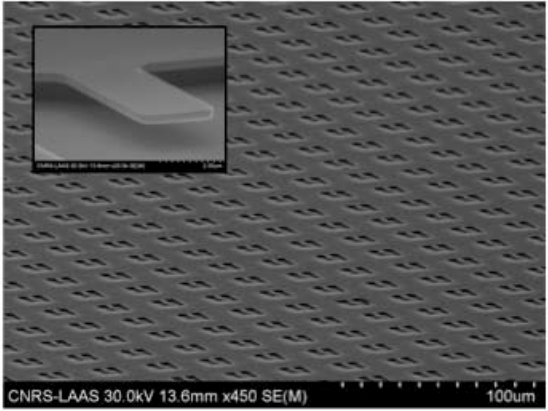

(a)

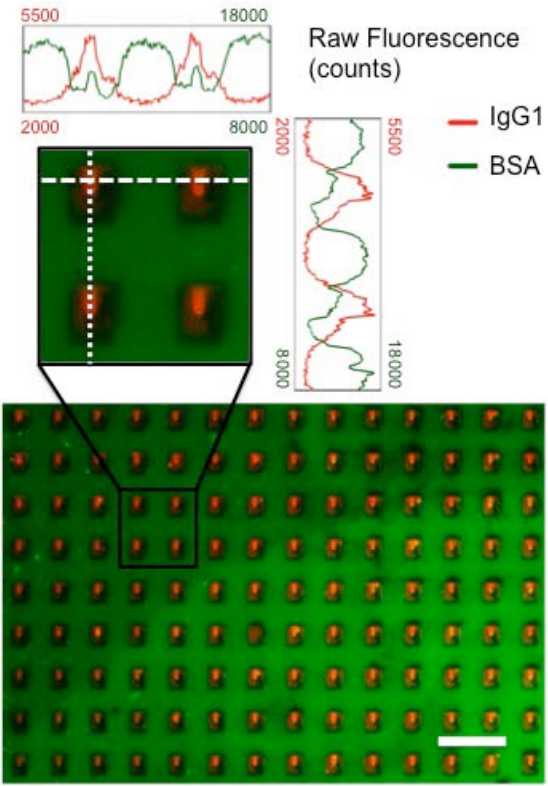

(b)

Figure 2. Biofunctionalized arrays of nanocantilevers. (a) Scanning electron microscopy images of fabricated silicon nanocantilevers. (b) Superimposed fluorescence pictures (taken with two dedicated filter sets) of an array of nanocantilevers functionalized after printing the Alexa Fluor 660 IgG1 and the Fluorescein-BSA in a single step (scale bar: 50µm). The inset provides a zoom of the array along with the fluorescence profiles measured with the two filters.

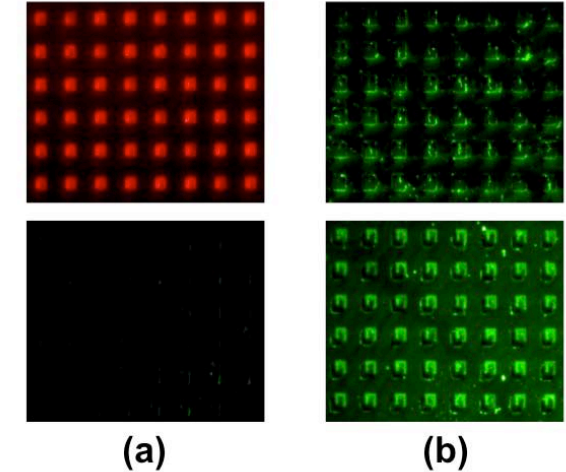

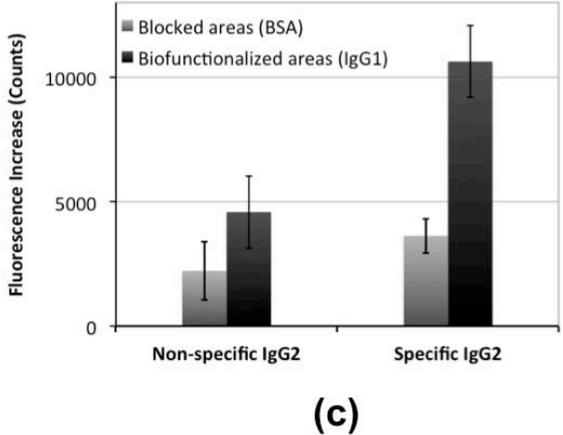

Figure 3. Validation of the bioactivity of the functionalized NEMS array using fluorescence microscopy: (a) Fluorescence images of the printed IgG1 using the filter sets respectively dedicated to observe the Alexa 660 and the Alexa 488 fluorophores (top and bottom), (b) Fluorescence images of biofunctionalized chips after incubation with the specific IgG2 and non-specific IgG2 (top and bottom, respectively), and (c) Graph comparing the mean increase of fluorescence signal due to specific and non-specific tagged IgG2 incubation, on the BSA layer and on the biofunctionalized cantilevers.

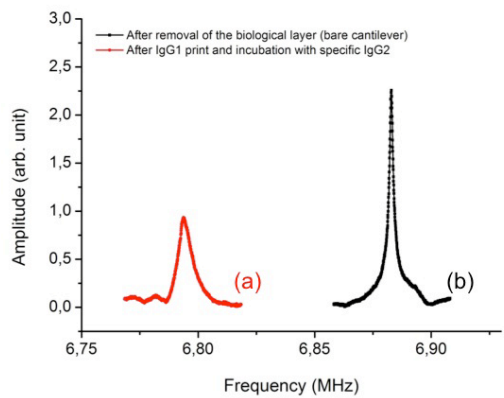

Figure 4. Resonant frequency curves of a single nanocantilever (a) after IgG1 print followed by the detection of specific IgG2, and (b) after removing all the printed and captured biomolecules.

| Mean resonant frequency shift [kHz] | Mean added mass [fg/µm$^2$] | Mean quality factor shift [%] |
|---|---|---|
| **-97,8** (15,5) | **18,6** (3) | **-66,2** (10,2) |

Table 1. Measured resonant frequency shift and quality factor drop induced by the functionalization (µCP) and passivation (batch) steps followed by biological recognition with the specific secondary IgG2. Frequency and quality factor shifts were calculated for 16 nanocantilevers of a single array and then averaged (the associated standard deviation is given within the brackets). The resulting calculated added mass is also provided in this table.

Supporting Information

Materials

A21083 Alexa Fluor 660 donkey anti-goat IgG (IgG1), A11001 Alexa Fluor 488 goat anti-mouse IgG (Specific IgG2), and A21202 Alexa Fluor 488 donkey anti-mouse IgG (Non-specific IgG2) were purchased from Invitrogen and diluted in Phosphate Buffered Saline (PBS) 1X solution at the respective concentrations of 400 μg/mL, 50 μg/mL, and 50 μg/mL. Albumin from bovine serum (BSA) and A23015 BSA fluorescein conjugate were purchased from Sigma and Invitrogen, respectively, and were both diluted at a final concentration of 1 mg/mL in 1X PBS.

Fabrication of NEMS arrays

Silicon on insulator (SOI) wafers were used as substrates for the fabrication of the nanocantilevers' arrays. From SOI wafers (340 nm thick P-type Si/1 μm $SiO_2$/525 μm thick Si, Soitec), cantilevers with controllable thickness can be fabricated and released in aqueous solution without structure sticking issues. The nanocantilevers were patterned using a UV stepper photo repeater (I Line CANON FPA 3000 i4/i5, N.A. 0.63) via a 600 nm thick positive photoresist layer (ECI). The top silicon layer was etched by reactive ion etching (RIE, Alcatel AMS4200) until the intermediate $SiO_2$ layer appeared. The sacrificial $SiO_2$ layer was then etched by dipping the wafer in a buffered hydrofluoric acid (BHF) solution leading to the release of the nanocantilevers. Finally, a 9 nm thick oxide layer was grown at the surface of the nanostructures by thermally oxidizing the chips in a dedicated furnace. Fabricated cantilevers are 8 μm long, 2.6 μm wide and 340 nm thick. Each 1.5 cm × 1.5 cm chip holds more than $10^5$ nanostructures.

Fabrication of PDMS stamps

The polydimethylsiloxane (PDMS) stamps were fabricated using Sylgard 184 products with a 1:10 ratio of curing agent and prepolymer. After mixing the two reagents, the mixture was carefully degassed. Standard 100 mm silicon wafers were used as master molds and the patterns were created using optical photolithography followed by a 340 nm deep reactive ion etching (DRIE) step. The master molds were silanized using OctadecylTrichloroSilane (OTS, Sigma, 1% in Xylene). The PDMS mixture was injected onto the master mold in a dedicated stainless steel set-up and cured at 60 °C for 4 h. Glass slides were placed in the set-up to create precise composite stamps consisting of patterned 1.5 cm × 1.5 cm × 400 μm PDMS films glued to the 1.5 cm × 1.5 cm × 1 mm thick glass slides (Figure S1).

Printing process

Freshly prepared PDMS stamps were first rinsed 10 times with 1 mL deionized (DI) water and then dried with a nitrogen gun. The stamps were inked by incubation during 2 min in a 400 μg/mL Alexa

Fluor 660 donkey anti-goat IgG 1X Phosphate Buffered Saline (PBS) solution. The stamps were rinsed 10 times with 1 mL PBS solution, 10 times with 1 mL with DI water, dried with nitrogen and then printed 6 times onto bare glass slides to remove all antibodies at the surface of the stamps, i.e. at the top of the trenches. To print the anti-fouling layer and the probe biomolecules in a single step (Figure 1B), stamps were additionally incubated for 2 min in a 1 mg/mL BSA solution, then rinsed and dried following the above-described procedure. Immediately after inking, the stamps were aligned with the oxygen plasma cleaned 1.5 cm × 1.5 cm NEMS chips using a microscope and an automated chuck (Karl Suss PA200 set-up). The alignment procedures were assisted by Moiré patterns to minimize alignment errors. The stamps and the chips were gently brought into contact and a backpressure was applied onto the stamps during 30 min in order to deform the patterns so that the bottom of the trenches could touch the nanocantilevers. The deposition of the anti-fouling layer in a batch process was carried out by incubating the NEMS chips in a degassed 1 mg/mL BSA solution for 30 min. The chips were then rinsed 10 times with 1 mL degassed PBS solution, 10 times with 1 mL degassed DI water and then dried with a nitrogen gun.

Molecular recognition and imaging

For biodetection purposes, the donkey anti-goat IgG functionalized NEMS chips were incubated in a 50 µg/mL Alexa Fluor 488 goat anti-mouse IgG solution or in a 50 µg/mL Alexa Fluor 488 donkey anti-mouse IgG solution for 2 h. After incubation, the NEMS chips were thoroughly washed 10 times with 1 mL degassed PBS solution. The chips were then soaked in a beaker containing 50 mL PBS solution and agitated with an orbital shaker (Heidolph Instruments Rotamax) at 50 rpm during 45 min. The samples were then rinsed 10 times with 1 mL degassed DI water and dried with a nitrogen gun. Complete removal of the biological layers from the nanocantilevers was achieved by immersing the NEMS chips into a Piranha bath during 10 min.

Fluorescence images were acquired using an Olympus IX70 inverted microscope, a 20X objective (N.A. 0.40, Olympus), two filter sets dedicated to the observation of Alexa Fluor 660 and Alexa Fluor 488 / Fluorescein dyes (U-M41008 from Chroma and U-MWIB3 from Olympus, respectively) and a Clara CCD ANDOR camera (DR-328G-C01-SIL, 5 s acquisition time).

Measurement of NEMS resonant frequency

The nanocantilever resonant frequencies were measured in vacuum using a Fabry-Perot interferometry bench. The NEMS chips were mounted on a piezoelectric disk, electrically actuated by a network analyzer (Agilent 4395A) and coupled to an in-house high-frequency amplifier. The device was placed inside a vacuum chamber pumped down to $10^{-6}$ mbar at room temperature. A 30 mW He-Ne laser (Melles Griot) was focused on the nanostructures using a beam expander and a long working distance microscope objective (20X, N.A. 0.28), leading to a 3 µm minimal beam size. Interferences were detected by a photodetector (New Focus 1601) connected to the network analyzer to track the response of single nanocantilevers at the excitation frequency.

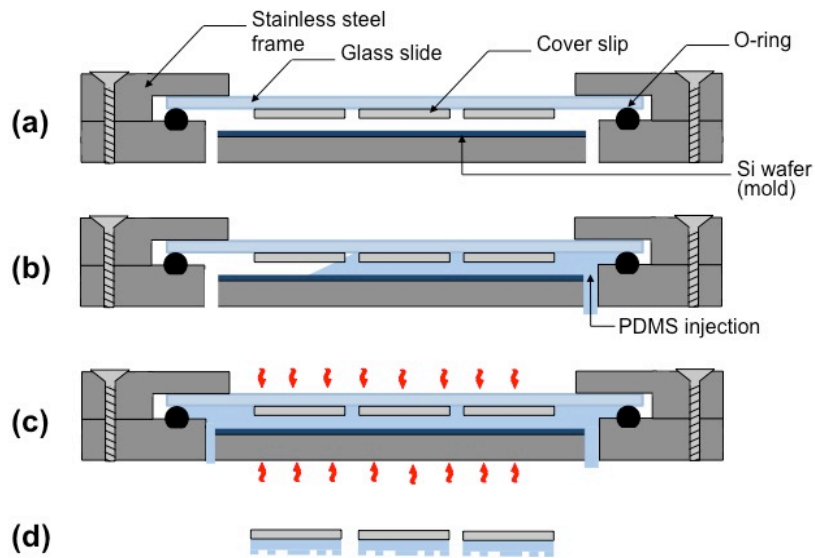

Figure S1. Schematics of the stamp fabrication: (a) Mounting the glass slide holding cover slips and the silicon wafer mold into the frame, (b) PDMS mixture injection, (c) PDMS curing, (d) Release of PDMS stamps stiffened by cover slips.

Figure S1. Schematics of the stamp fabrication using a dedicated stainless steel set-up.

Movie M1. Real-time video of the printing process. After bringing the stamp into contact with one side of the NEMS chip, the PDMS adhesion front moves along the chip, demonstrating the appropriate contact of the PDMS stamp with the chip. In a second step, a pressure is applied onto the stamp to collapse the bottom of the patterned trenches onto the nanocantilever for biofunctionalization purpose.